\documentclass[twocolumn,preprintnumbers,
endnote,nofootinbib,prl]{revtex4}
\usepackage{graphicx}
\usepackage{color}

\usepackage{amsfonts}

\newcommand{\vev}[1]{\langle {#1} \rangle}
\newcommand{\lsim}{\lesssim}
\newcommand{\gsim}{\gtrsim}

\newcommand{\eq}[1]{Eq.~(\ref{#1})}

\newcommand{\ord}[1]{\mathcal{O}{(#1)}}
\newcommand{\beq}{\begin{equation}}
\newcommand{\eeq}{\end{equation}}
\newcommand{\bea}{\begin{eqnarray}}
\newcommand{\eea}{\end{eqnarray}}

\begin{document}

\pagestyle{plain}

\title{\boldmath Variation of $\alpha$ from a Dark Matter Force}

\author{Hooman Davoudiasl
\footnote{email: hooman@bnl.gov}
}

\author{Pier Paolo Giardino
\footnote{email: pgiardino@bnl.gov}
}

\affiliation{Physics Department, Brookhaven National Laboratory,
Upton, NY 11973, USA}


\begin{abstract}

We consider a long range scalar force that mainly couples to dark matter and unstable Standard Model states, like the muon, with tiny strength.  Probing this type of force would present a challenge 
to observations.  We point out that the dependence of the induced background scalar field on dark matter number density can cause the mass of the unstable particles to have spatial 
and temporal variations.  These variations, in turn, leave an imprint on the value of the fine structure constant $\alpha$, through threshold corrections, that could be detected in astronomical and cosmological measurements.  Our mechanism can accommodate the mild preference of the Planck data for such a deviation, $(\alpha_{_{\rm CMB}}-\alpha_{\rm present})/\alpha_{\rm present} = (-3.6\pm 3.7)\times 10^{-3}$.  In this case, the requisite parameters typically imply that violations of Equivalence Principle may be within reach of future experiments.

\end{abstract} \maketitle

\section{Introduction}
Though dark matter (DM) makes up about a quarter of the energy budget in the Universe, its 
properties remain mostly unknown \cite{Patrignani:2016xqp}.  In particular it is not known whether DM has any long range interactions other than gravity.  If such a ``dark'' force exists, it could affect the long distance 
dynamics of DM, potentially providing a better understanding of the observed large scale structure. In any event, given the existing data, such interactions must be quite weak; if they extend over galactic scales, likely they are not allowed to be much stronger than gravity.

Once one accepts that DM may have long range interactions, it is natural to ask what other 
states are coupled to such a force.  If the particles in question are the stable constituents of 
atoms, the electron and nucleons, the strength of their coupling to the long range force is extremely 
well constrained by tests of the Equivalence Principle and ``fifth force'' searches, requiring the 
strength of those interactions to be sub-gravitational.  This situation could limit the effects of the new interactions, though there are potentially interesting scenarios that can arise in this case \cite{Davoudiasl:2017pwe}.  However, one could also entertain the 
possibility that the long range interactions of DM couple more strongly to other more elusive  
Standard Model (SM) particles, like neutrinos \cite{Davoudiasl:2018hjw} or unstable particles, 
such as the muon.  In the latter case, the absence of these particles on macroscopic scales 
does not allow very stringent experimental constraints on their new long range interactions.  For the 
same reason, it seems quite challenging to envision how one may uncover a new long distance 
force between unstable particles and DM.

In this work, we consider the coupling of a long range force, mediated by a light scalar $\phi$ 
to DM and an {\it electrically charged} unstable SM fermion $f$; for concreteness we will focus on the muon.  
We show that the the background field $\phi$ sourced by the 
cosmic population of DM can result in variations 
of the fermion mass $m_f$, in space and time, which leaves its imprint as a threshold effect 
in the running of fine structure constant $\alpha$ of quantum electrodynamics. The possibility that fundamental constants may vary has been considered in previous works, starting from Dirac's large numbers hypothesis \cite{Dirac:1937ti}, see for example Refs.~\cite{Terazawa:1981ga, Bekenstein:1982eu, Marciano:1983wy, Stadnik:2015kia} and references therein. However, contrary to most previous models, in our case the variation of $\alpha$ is tied to the local density of DM and is not  simply correlated with the evolution of the Universe.

Scalar long range forces may be motivated from top down or phenomenological points of view \cite{Friedman:1991dj, Gradwohl:1992ue, Dolgov:1999gk, Farrar:2003uw, Gubser:2004du, Gubser:2004uh, Nusser:2004qu, Kesden:2006vz, Kesden:2006zb, Farrar:2006tb}.  In general, one has to ensure that the mass of $\phi$ stays small under quantum corrections, and also that its renormalized potential is sufficiently small, since the interaction of $\phi$ with its own background field would generate a potentially large mass term.  This issue is a generic feature of the models that require the existence of a long range ``fifth force" mediated by a scalar, and its solution may be found in supersymmetry or string dynamics \cite{Gubser:2004du}.

In the current understanding  of quantum field theory (QFT), without invoking special symmetries, scalars that are light compared to other scales of a theory  
require a commensurate degree of fine-tuning.  In this paper, we are interested in the phenomenological effects of this force, and we will not comment further on its naturalness, noting only that the discovery of such a field 
would likely require a revision of the presently accepted views on QFT.

\section{Long Range Force}
In this section, we describe our mechanism.
The basic interactions of interest for our analysis are given by 
\beq
{\cal L}_i = - g_X \phi \bar X X - g_\mu \phi \bar \mu \mu \,,
\label{Li}
\eeq
where $X$, a Dirac fermion, is the DM, $\phi$ is a light scalar that mediates the long-range force, and $\mu$ is the SM muon.
In general, other SM fermions could enter in Eq.~(\ref{Li}), but we found the choice of the muon particularly interesting and we will concentrate on it for the rest of this letter. 
We will assume that the effective dimension-4 operator $\phi \bar \mu \mu$ is the low energy result of some well-behaved but un-known UV theory.
The relevant mass terms, {\it in vacuo}, are given by 
\beq
{\cal L}_m = -m_X \bar X X - m_\mu \bar \mu \mu - \frac{1}{2} m_\phi^2\, \phi^2\,,
\label{Lm}
\eeq
in an obvious notation. 

Let us consider what happens when a sufficiently large density of DM $X$ fermions is present.  The 
equation of motion for $\phi$ is then given by (see, for example, Ref.~\cite{Gubser:2004du})
\beq
(\Box + m_\phi^2) \phi = - g_X \bar X X = - g_X n_X \vev{\sqrt{1 - v^2}} {\rm sgn}(\phi),
\label{Boxphi}
\eeq
where $n_X$ is the number density of $X$, $\vev{\dots}$ denotes an average, 
and $v$ is the velocity of $X$.  
Here, we assume that the population of $\mu$ states is 
negligible.  The second equation 
contains a factor $\sqrt{1-v^2}$, since $\bar X X$ is Lorentz invariant.  We are interested in DM well after its relic density has been set, and hence 
we can assume $v\approx 0$.  

If the distribution of DM is static and uniform, and has a 
characteristic size that is larger than all other distance scales of interest $\Box \phi \approx 0$, hence 
\beq
\phi \approx - \frac{g_X n_X}{m_\phi^2}\,.
\label{phi}
\eeq

According to \eq{Li}, the contributions to the mass of $\mu$ and $X$ 
from the scalar force are given by 
\beq
\Delta m_F = g_F\, \phi\,,
\label{DelmF}
\eeq
where $F=\mu, X$. An interesting consequence of the modification of the mass of the muon is that, due to threshold corrections,\footnote{For other works in different contexts see, for example,  Refs.~\cite{Chacko:2002mf, Dent:2003dk}.} the fine structure constant $\alpha$ changes as well according to 
\beq
\frac{\Delta\alpha}{\alpha}=\frac{2\alpha}{3\pi}\ln(1+\frac{\Delta m_\mu}{m_\mu}).
\eeq

\begin{figure}
\includegraphics[width=0.35\textwidth]{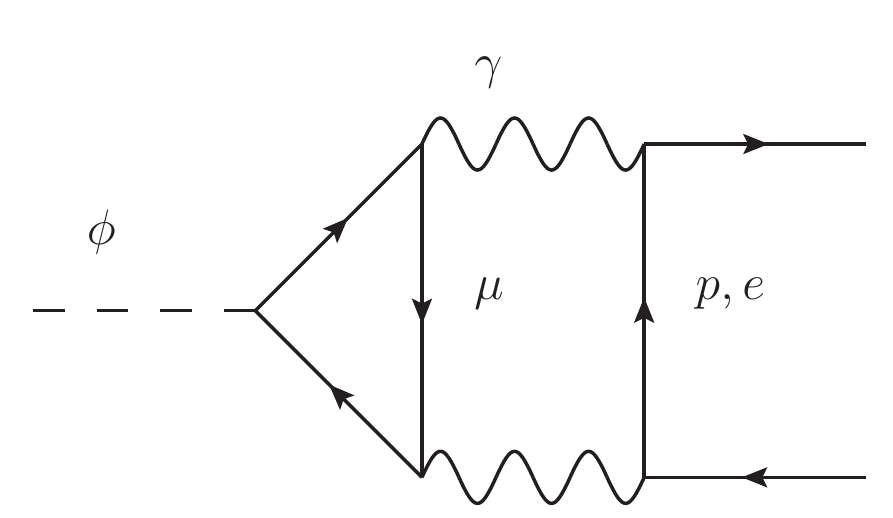}
 \caption{Loop-induced coupling between $\phi$ and a proton ($p$) or electron ($e$)}
 \label{Fig1}
\end{figure}

The coupling of $\phi$ to the muon typically implies tiny couplings between $\phi$ and other SM fermions, via radiative corrections. In particular, the diagram in Fig.~\ref{Fig1} contributes to the coupling between $\phi$ and stable fermions (electron and proton, respectively $g_e$ and $g_p$) even if they are zero in the tree-level Lagrangian\footnote{We thank W. Marciano for pointing out the potential significance of these  diagrams.}.
The presence of this long range interaction is observable in tests of the Equivalence Principle, see for example \cite{Bovy:2008gh, Carroll:2008ub, Carroll:2009dw}, thus providing indirect bounds on the $g_\mu$ coupling. From Refs.~\cite{Berge:2017ovy,Fayet:2017pdp}, we find that 
$|g_p|\lesssim10^{-24}$ and $|g_e| \lsim 10^{-25}$. At the same time, given the 2-loop diagram in Fig.~\ref{Fig1} we would expect the coupling to protons to be\footnote{Note that, as in other similar work cited here, the fine-tuning of quantum corrections required for a small value of $m_\phi$ does not imply that these 2-loop diagrams are also tuned away.  Thus, our estimates of their size are consistent with the phenomenological assumptions about the theory.}
\beq
g_p\sim \frac{\alpha^2}{(4\pi)^2} \frac{m_\mu}{m_p} g_\mu,
\label{gp}
\eeq
corresponding to an upper bound $|g_\mu|\lesssim 10^{-17}$.  Notice that this would ensure for  the electron coupling $|g_e|\lesssim10^{-25}$, due to an $\ord{m_e/m_\mu}$ suppression.
Limits on the $g_X$ coupling are less strict. It is reasonable to require $g_X$ to be (sub-)gravitational if the range of the force is of galactic scale, in order to avoid conflict with our present understanding of large scale structures. On the other hand, we can relax this requirement if we consider smaller ranges ($i.e.$ heavier mass) for $\phi$.

\section{Consequences}
We now consider a particular scenario where the range of the force mediated by $\phi$ is $100$ kpc, so that it spans the Milky Way and the majority of its halo; $m_\phi = 1/100$~kpc $\sim 10^{-28}$ eV. We set the mass of DM $m_X=1$ GeV, for concreteness, and since we require the force mediated by $\phi$ to be sub-gravitational this corresponds to imposing $|g_X|\lesssim 10^{-19}$; thus we fix $g_X=5\times10^{-20}$. Notice that since $g_X\sim 10^5 g_p$ the contribution of common matter to the value of $\phi$ is negligible and Eq.~(\ref{Boxphi}) is valid. We note that the form of Eq.~(\ref{Boxphi}) suggests that if we scale $g_X$ proportional to the DM mass, that is for constant  ``gravitational charge," the underlying physics stays the same, since $n_X \propto 1/m_X$. We also set the coupling to the muons at $g_\mu=-2\times10^{-18}$ so that the contribution to its mass is positive, as implied by \eq{DelmF}. Later, we will also consider an interesting case with $g_\mu >0$.

For the DM distribution in the Milky Way we consider the NFW and Burkert profiles \cite{Navarro:1995iw,Burkert:1995yz}, respectively
\beq
\rho_{\text{NFW}}=\frac{\rho_n}{(r/R)(1+r/R)^2}
\eeq
and
\beq
\rho_{\text{Burkert}}=\frac{\rho_b}{(1+r/r_c)(1+(r/r_c)^2)}\,,
\eeq
where we took $R=20$ kpc and $r_c=10$ kpc. Here, $\rho_n$ and $\rho_b$ are chosen so that the local density of DM in the solar system ($r=8.5$ kpc) is $0.3$ GeV/cm$^3$. We assume a spherical distribution.

\begin{figure}
\includegraphics[width=0.45\textwidth]{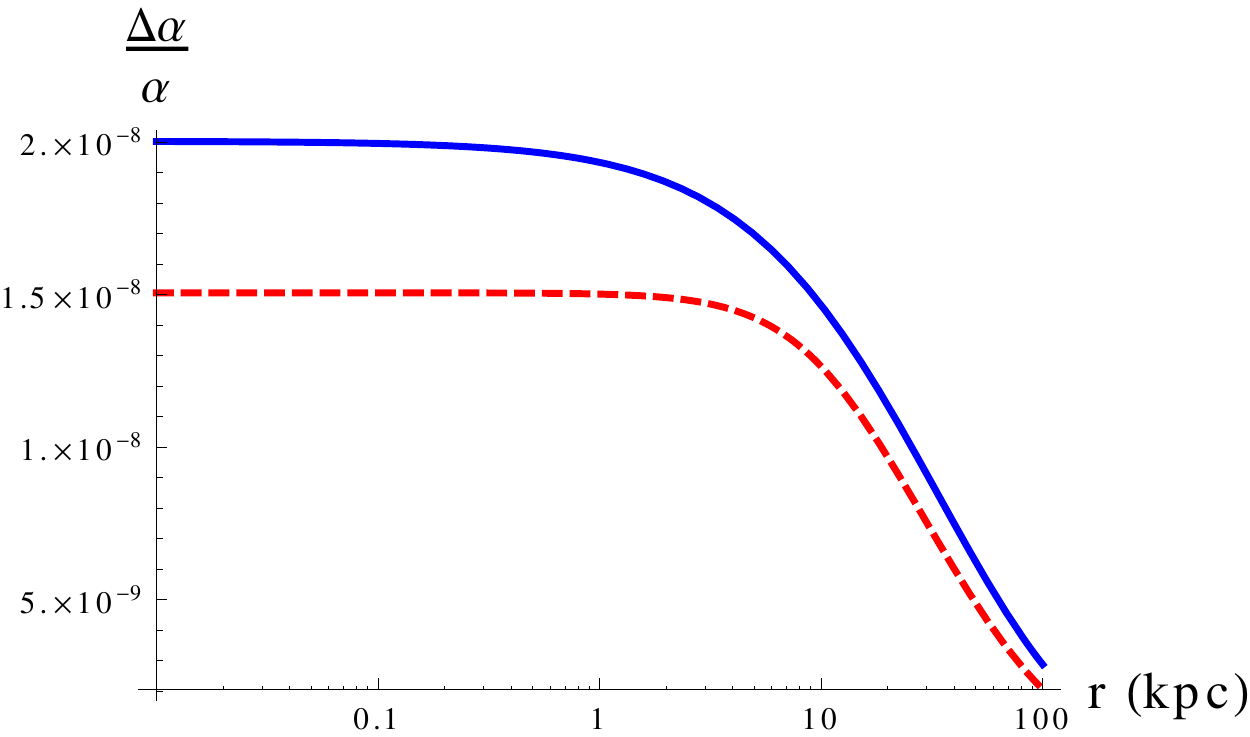}
 \caption{$\Delta\alpha/\alpha$, using the first set of benchmark parameters, $m_\phi=1/100$ kpc$^{-1}$,  $g_\mu=-2\times10^{-18}$ and $g_X=5\times10^{-20}$,  as a function of the distance from the center of the Galaxy in kpc. The blue solid line is obtained assuming the NFW distribution for DM. The red dashed line assumes the Burkert profile.}
 \label{Fig2}
\end{figure}

We solved \eq{Boxphi} numerically, assuming $\partial_r\phi|_{r=0}=\phi(\infty)=0$ and the above DM profiles, and obtained the value of $\Delta\alpha/\alpha$ as a function of distance from the center of the Galaxy. Here, the variation is with respect to the value in vacuum: 
$\Delta \alpha \equiv \alpha - \alpha_{\rm vac}$.  
In Fig.~\ref{Fig2},  we plot our results. We consider particularly interesting the fact that the value of $\Delta\alpha/\alpha$ at the center of the Milky Way is $\mathcal{O}(10)$ times larger than its value at the outskirts. 

\begin{figure}
\includegraphics[width=0.45\textwidth]{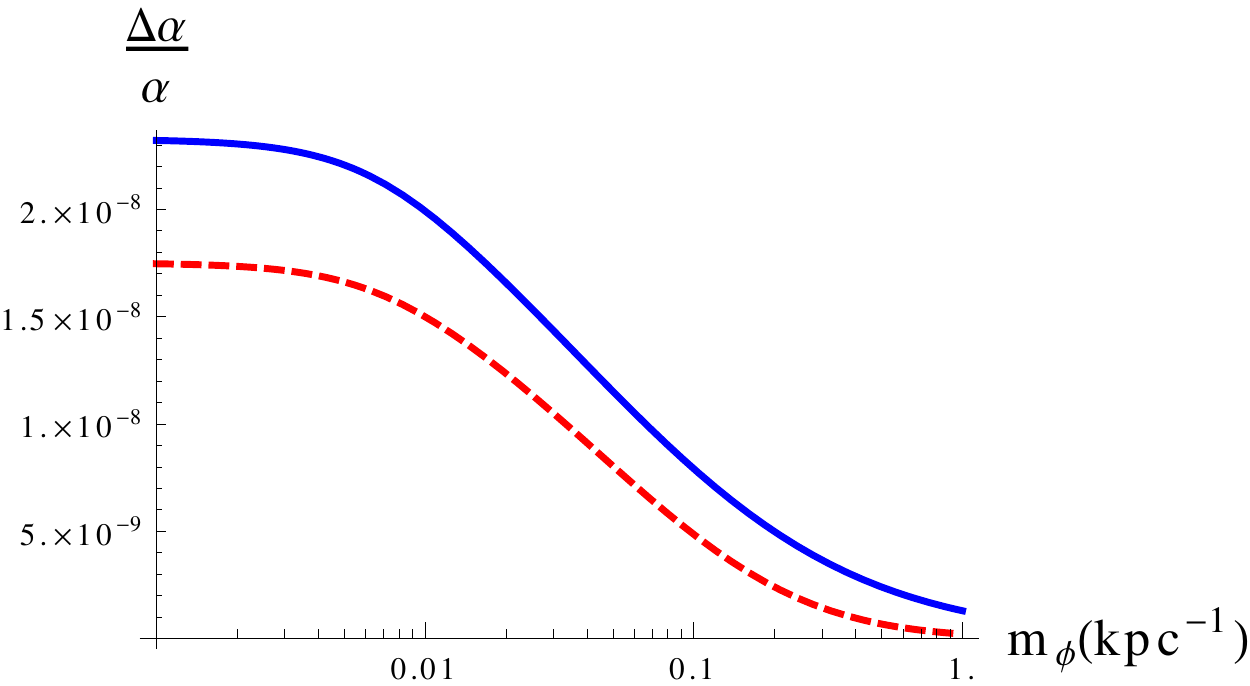}
 \caption{$\Delta\alpha/\alpha$ at the center of the Galaxy versus $m_\phi$, for NFW (solid blue) and Burkert (dashed red) choices of DM profile.  Here, $g_X=  5(m_X/{\rm GeV})\times 10^{-20}$ has been assumed.}
 \label{Fig3}
\end{figure}

In Fig.~\ref{Fig3} we plot the value of $\Delta\alpha/\alpha$ at the center of the Galaxy, for values of $m_\phi$ between 0.001 and 1 kpc$^{-1}$. For simplicity we set $g_X=  5(m_X/{\rm GeV})\times 10^{-20}$, so that the result does not depend on the mass of DM. As we can see, for a wide range of values of $m_\phi$, the typical change in $\alpha$ is $\sim 10^{-8}$, for both the NFW (solid blue) and Burkert (dashed red) choices of DM profile.

Focusing on the solar system, we find that  the mass of the $\mu$ lepton receives a contribution due to DM in the Milky Way of $\Delta m_\mu/m_\mu\sim 10^{-5}$, that corresponds to a variation of $\alpha$ from its value {\it in vacuo}:
\beq
\frac{\Delta\alpha}{\alpha}\sim10^{-8}.
\eeq
While $\Delta m_\mu$ corresponds to a deviation of the SM muon Yukawa too small to be accessible at the LHC, $\Delta\alpha/\alpha$ is close to the present bounds obtained from the Oklo natural reactor:
\beq
\frac{\Delta\alpha}{\alpha}\sim10^{-8}\text{\textemdash}\,10^{-7};
\eeq
see for example Ref.~\cite{Petrov:2005pu, Gould:2007au, Onegin:2010kq} and references therein.

The above result can be interpreted in our scenario as a constraint on how much the density of DM of our Galaxy changed in the course of the last 2 billions years, since the activity period of the Oklo reactor. So we can conclude that, in our scenario, variations of order $\mathcal{O}(1)$ in the overall mass density  of the Milky Way halo are allowed.  
This is likely much more than the amount of DM accreted through 
mergers with the satellites of the Milky Way.  On the other hand, the above results imply that an $\ord{10}$ more stringent constraint from Oklo or other similar measurements can be sensitive to $\sim 10\%$ DM accretion by the Milky Way, over time scales of $\ord{10^9}$ years.

Measurements of  $\alpha$  in other galaxies are usually less constraining \cite{Uzan:2010pm, Martins:2017yxk} and the current bounds are generally of order $\sim10^{-6}$ for $\Delta\alpha/\alpha$, that would easily accommodate a few orders of magnitude of difference in the density of DM among various galaxies.  

Another interesting consequence of this scenario is that the values of the muon mass and $\alpha$ depend on the {\it cosmological era}. Since the density of DM is proportional to the cube of the temperature of the Universe, if we go back in time ($i.e.$ at higher temperatures) we expect the mass of the muon and the value of $\alpha$ to change. However, the horizon size, $d_{\text{hor}}$, also depends on the temperature of the Universe and shrinks as we go towards earlier times. Thus, we would eventually reach a point in time where $d_{\text{hor}}<m_\phi^{-1}$ is the meaningful scale in the calculation of $\phi$.  We have, up to $\ord{1}$ corrections, 
\beq
\phi \sim  - g_X n_X d_{\text{hor}}^2\propto \left\{
\begin{array}{ccc}
\text{consant}& ; & \text{matter-dominated} \\
\frac1{T}& ; & \text{radiation-dominated}   
\end{array}
\right.,
\eeq
where $n_X\propto T^3$ and $d_{\text{hor}}\propto 1/H\,\text{or}\,2/H$ if the Universe is either radiation or matter dominated, respectively.  Here, $H$ denotes the Hubble scale.  In what follows, we will assume the Universe is dominated by matter or radiation when the corresponding energy density dominates by a factor of 10.  In between these two regimes, we use a simple linear function to interpolate between $2/H$ and $1/H$.  As $H$ grows with $T$, $\phi$ eventually decreases. 

\begin{figure}
\includegraphics[width=0.45\textwidth]{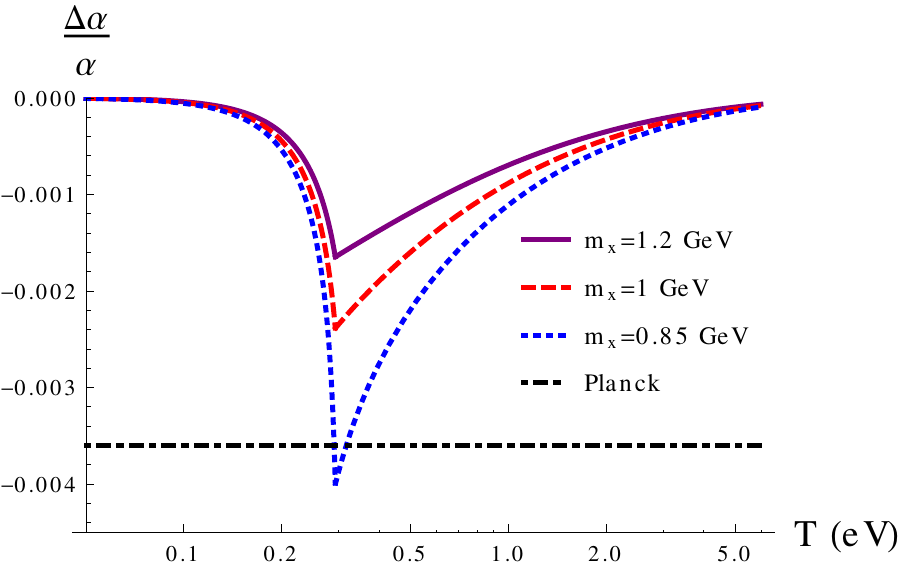}
 \caption{$\Delta\alpha/\alpha$, using the second set of benchmark parameters, $m_\phi=1/300$ kpc$^{-1}$,  $g_\mu=10^{-18}$ and $g_X=2\times10^{-21}$,  as a function of the temperature  ($T$) of the Universe, for three values of $m_X$.  The central value of the Planck result $\Delta\alpha/\alpha=(-3.6\pm3.7)\times10^{-3}$ \cite{Ade:2014zfo} is marked by the dot-dashed line.}
 \label{Fig4}
\end{figure}

In the scenario that we explore here $\phi$ reaches its maximum at $T\sim 1$ eV at which point $\Delta m_\mu \sim 600$ MeV. This large value for the mass of the muon is not problematic by itself, since at those temperatures muons are out of equilibrium and do not play a role in cosmological evolution anymore. Also $\Delta m_\mu$ is large only in a small window around $T\sim 1$ eV  and $\Delta m_\mu/m_\mu \ll10^{-3}$ during the Big Bang Nucleosynthesis and earlier epochs. However such a large value of the $\mu$ lepton mass affects the fine structure constant and we have, for $T\sim 0.3-1$ eV
\beq
\frac{\Delta\alpha}{\alpha}\sim (2\text{\textemdash}\,5)\times 10^{-3}.
\eeq
This result is particularly interesting if we consider that the Planck experiment \cite{Ade:2014zfo} found a difference\footnote{See also Ref.~\cite{Hart:2017ndk}.} between the value of $\alpha$ at the CMB era with respect to today's measurement of $\Delta\alpha/\alpha=(-3.6\pm3.7)\times10^{-3}$ (note that our convention for $\Delta \alpha$ differs by a minus sign from that of Ref.~\cite{Ade:2014zfo}). Our benchmark parameters are compatible with this measurement, within $2\sigma$.

Alternatively, one could assume the central value of the above Planck result to furnish a mild indication 
that $\Delta \alpha/\alpha \sim -10^3$ is preferred.  This can be achieved in our scenario by modifying the benchmark parameters of our model. Taking $m_\phi=1/300$ kpc$^{-1}$,  $g_\mu=10^{-18}$ and $g_X=2\times10^{-21}$ we obtain $m_\mu\sim 20 $ MeV at  $T=0.3$ eV. In Fig.~\ref{Fig4}, we plot $\Delta\alpha/\alpha$ as a function of temperature for three values of $m_X=0.85,1.0, 1.2$~GeV.  As one can see, our model can accommodate the central value of the Planck measurement, for $m_X\lesssim 1$ GeV.  Whether or not this mild hint will grow in significance, our results  point to the possibility of constraining DM long-range interactions through measurements of the variations of physical constants in different eras. Notice also that for a larger $g_X$ the muon could become lighter than the electron for a short period before and after CMB, which would allow the electron to decay into a muon and 
neutrinos! This would have unusual  
effects on cosmology that we will not further consider in this letter.  Here, we add that for the first and second sets of benchmark parameters considered above the DM mass does not vary by more than $\sim 10^{-2}$ and $10^{-4}$, respectively, which are allowed by the current  percent level determinations of the DM energy density \cite{Patrignani:2016xqp}. 

If the central value of the Planck measurement for $\Delta \alpha/\alpha$ holds near 
its current value with improved measurements, the scenario discussed above could typically 
imply violations of the Equivalence Principle, not far from the current limits.  
To see this, note that increasing $g_X$ by more than an order   
of magnitude will lead to conflict with the CMB measurements of the DM energy density, as this would change $m_X$ more than $\sim 1\%$ for $T\sim1$~eV.  Therefore, to stay near the Planck central value we need $g_\mu \gsim 10^{-19}$.  Then, \eq{gp} implies that $g_p \gsim 10^{-26}$, which is within two oder of magnitudes of the current limits.

Lastly, let us mention that the large positive change in the mass of the muon around CMB era can have  another interesting consequence for light thermal relic DM. If the DM thermal relic density is dominantly set through the annihilation into $\mu^+ \mu^-$ final states\footnote{Or other exotic fermions that change their masses dramatically during the CMB era through the mechanism described here.}, the process could be allowed in early and late cosmology, but become forbidden during the CMB era, thus relaxing the current  bounds \cite{Madhavacheril:2013cna, Ade:2015xua} on the thermal relic abundance of light DM.

\section{Concluding Remarks}
In this work, we have examined a possible signal of a long range force, 
mediated by a light scalar, that couples to DM with order gravitational 
strength, but could have somewhat larger couplings to unstable SM particles.  Given 
the feebleness of the assumed interactions and the lack of significant 
populations of the unstable states, this scenario can pose a significant 
challenge to experimental verification.  We show that if the SM 
particles have electric charge, the scalar potential sourced by DM can modify the 
threshold effects in the
running of the fine-structure constant $\alpha$ and lead to its variations 
in space and time, as a function of DM density.  Focusing on the muon for 
concreteness, we found that for phenomenologically allowed values of 
parameters existing bounds on variations of
$\alpha$ can be satisfied.   

In the early Universe,  when the density of 
DM was much larger, we expect sizable  deviations in $\alpha$, however our 
benchmark parameters are consistent with the current Planck bound from 
the CMB era.  Depending on the sign of the Yukawa couplings to the 
mediating scalar, one could realize
a positive or negative deviation; the latter choice is modestly preferred 
by the Planck data and can be accommodated by our scenario.  We 
conclude that future improvements in these or other astrophysical data can 
potentially uncover the effect of the long range scalar force on $\alpha$. 
If the Planck hint holds, our mechanism typically predicts 
violation of the Equivalence Principle, not far from 
present bounds.
Our proposal hence provides a handle on an otherwise extremely elusive 
possible phenomenon, whose discovery would have a revolutionary impact on 
our understanding of particle physics and cosmology.

\section*{Acknowledgements}
We thank W. Marciano, G. Mohlabeng, N. Sehgal, and M. Sullivan for comments and discussions.  This work is supported by the U.S. Department of
Energy under Grant Contracts DE-SC0012704.


\bibliography{LR-DM-alpha}


\end{document}